\documentclass[aps,pre,twocolumn,final,floatfix,showpacs,amsmath,amssymb,10pt]{revtex4}

\newcommand*{\Base}[1][i]{\ensuremath{r_{#1}}}
\newcommand*{\Sequence}{\ensuremath{\mathcal{R}}}
\newcommand*{\Structure}{\ensuremath{\mathcal{S}}}
\newcommand*{\epsStructure}{\ensuremath{\Structure_\varepsilon}}
\newcommand*{\SM}[1][i,j]{\ensuremath{S_{#1}}}
\newcommand*{\Ensemble}{\ensuremath{\mathcal{E}}}
\newcommand*{\PairE}{\ensuremath{E_p}} % pair energy
\newcommand*{\StackE}{\ensuremath{E_s}}% stacking energy
\newcommand*{\randE}{\ensuremath{\eta}}% random field energy
\newcommand*{\N}{\ensuremath{N}}
\newcommand*{\Nh}{\ensuremath{\hat N}}
\newcommand*{\Z}{\ensuremath{Z}}
\newcommand*{\Zh}{\ensuremath{\hat Z}}
\newcommand*{\kb}{\ensuremath{k_\mathrm{B}}}
\newcommand*{\kbT}{\ensuremath{k_\mathrm{B}T}}
\newcommand*{\overlap}{\ensuremath{q}}
\newcommand*{\distance}{\ensuremath{d}}
\newcommand*{\Ssize}{\ensuremath{t}} % average stacking size
\newcommand*{\BC}{\ensuremath{B}} % Binder Cumulant
\newcommand*{\PC}{\ensuremath{A}} % Parisi Cumulant
                                
\newcommand*{\average}[1]{\ensuremath{\left\langle #1\right\rangle}}
\newcommand*{\qaverage}[1]{\ensuremath{\left[ #1\right]}}

\newcommand*{\define}{:=}
\newcommand*{\E}{\mathrm{e}}

\newcommand*{\order}[1]{\ensuremath{\mathcal{O}(#1)}}

\newcommand*{\qvariance}{\chi_{\Sequence}}

\usepackage{xspace}
\newcommand*{\eg}{e.g.,\xspace} % z.B.
\newcommand*{\ie}{i.e.,\xspace} % d.h.
\newcommand*{\refeq}[1]{eq.\@ \eqref{#1}\xspace}
\newcommand*{\reffig}[1]{Fig.\@ \ref{#1}\xspace}
\newcommand*{\reftab}[1]{Tab.\@ \ref{#1}\xspace}
\newcommand*{\refsec}[1]{section\@ \ref{#1}\xspace}
\newcommand*\epsCoup{$\varepsilon$-coupling\xspace}

\renewcommand\atop[2]{\genfrac{}{}{0pt}{}{#1}{#2}}

\usepackage{graphicx}
\newcommand*\picturewidth{1.0\columnwidth}

\newcommand*\guidelines{Calculated data points are indicated by symbols. Lines
  are  drawn to guide the eye.} 

\newcommand*\errorbars{Missing error bars are of the size of the symbols or smaller,  and omitted for legibility.}

\newcommand*\legendsymbols{%
\label{legendsymbols}% use only once
For system size $L$ the following symbols are used: 
$\circ$~128, $\triangle$~192, $\square$~256, $\triangledown$~384, 
$\Diamond$~512, $\triangleleft$~768, $\blacksquare$~1024.
\guidelines\ \errorbars
}
\newcommand*\legendref{For explanation of symbols see caption of
  \reffig{legendsymbols}.\xspace}

\begin{document}

\author{Bernd Burghardt}
\email{burghard@theorie.physik.uni-goettingen.de}
\author{Alexander K. Hartmann}
\email{hartmann@theorie.physik.uni-goettingen.de}
\affiliation{Institut f{\"u}r Theoretische Physik, Universit{\"a}t G{\"o}ttingen,
  Friedrich-Hund-Platz 1, D--37077 G{\"o}ttingen, Germany}

\title{Dependence of RNA secondary structure on the energy model}
\date{\today}
\begin{abstract}
We analyze a microscopic RNA model, which includes two widely used models as
limiting cases,   namely it contains terms for bond as
  well as for stacking energies.
We numerically investigate possible changes in the qualitative and quantitative
behaviour while going from one model to the other; 
in particular we test, whether a transition occurs, when continuously
moving from one model to the other. 
For this we calculate various thermodynamic quantities, both at zero
temperature as well as at finite temperatures.
   All calculations can be done efficiently in polynomial time by a dynamic
programming algorithm.  
We do not find a sign for  transition between the models, 
but the critical
exponent $\nu$ of the correlation length, 
describing the phase transition in all models to an
ordered low-temperature phase, seems to depend continuously 
on the model. Finally,
we apply the \epsCoup method, to study low excitations. The
exponent $\theta$ describing the energy-scaling of the excitations
seems to depend not much on the energy model. 
  \end{abstract}

\pacs{64.60.Fr, 87.15.Aa}

\maketitle
%%%%%%%%%%%%%%%

\section{Introduction}
RNA plays an important rule in the biochemistry of all living
systems~\cite{GCA99,Hig00}.  Similar to the DNA, it is a linear chain-molecule
build from four types of bases, \ie adenine (A), cytosine (C), guanine (G),
and uracil (U).  
It does not only transmit pure genetic information, but, \eg works as a
catalyst. While for the former the primary structure, \ie the sequence of
the bases, is relevant, for the later the kind of higher order structures, \ie
secondary and tertiary structures, are relevant. 

%%%% hydrogen bonds
Like in the double helix of the DNA, in RNA complementary bases can build
hydrogen bonds between each other. Compared to DNA, where the bonds are built
between two different strands, RNA builds bonds between bases of the same RNA
strand. The information, which bases of the strand are paired, gives the
secondary structure, and the spatial structure is called the tertiary
structure. 
The tertiary structure is stabilized by a much weaker interaction than the
secondary structure. This leads to a separation of energy scales between
secondary and tertiary structure, and gives the justification  to neglect the
later~\cite{BH99}. 
Therefore we deal here with the secondary structure only. 

One crucial point in calculating the secondary structure is the used energy
model: On the one hand, if one aims to get minimum structures close to the
experimentally observed one, one uses energy models that take into account
structure elements~\cite{Zuk89,McC90,HFS*94,LZP99}, \eg hair pin loops. 
On the other hand, if one is interested in the qualitative behaviour, one uses
models as simple as possible that keep the general behaviour, \eg only one
kind of base~\cite{LB04} or using energies depending only on the number and on
the type of paired bases~\cite{Hig96,BH02,MPR02,PPR00}. 
Here we will consider only models with the later kind of interaction energy. 
In recent years several authors examined this kind of models with regard to
the thermodynamic behaviour, \ie searching for phase transitions and
describing the type of phase involved~\cite{BH02,PPR00,Har01,MPR02}; Liu and
Bundschuh~\cite{LB03,LB04} recently discussed, whether  native RNA is already
in the regime of the thermodynamic limit or finite size effect have to be
taken into account. 
In this paper we numerically investigate a hybrid model of two well known
energy models \cite{HB03,MET*03},
 \ie a \emph{pair energy model}, where only base pairs are
considered regardless of their neighbourhood, and a \emph{stacking energy
  model}, where only consecutive paired bases, \ie forming a stack, gives an
energy contribution. 
 It has been claimed 
that the stacking energy is more relevant than just
the pair energy in real RNA \cite{TB99}.
Our model contains terms for {\em both} types of
  interactions and allows to move continuously from one model to the
  other. We are interested, whether the two limiting models are
  qualitatively different, in particular, whether a phase transition
  occurs, when moving from one model to the other.

The paper is organized as follows. In section \refsec{sec:model}, we define
our model, \ie we formally define secondary structures and introduce our energy
model. 
In \refsec{sec:algorithm}, we explain how to calculate the partition function
with a dynamic programming algorithm. 
In \refsec{sec:observables}, we introduce the observables which we
investigate in the following \refsec{sec:results}. 
While in \refsec{sec:binder_cumulant} and \refsec{sec:temperature_dependence}
we do  finite temperature calculations, in \refsec{sec:epsilon_coupling}
we use the \epsCoup method at zero temperature.

%%%%%%%%%%%%%%%%%%%%%%%%%%%%%%%%%%%%%%%%%%%%
\section{The Model}
%%%%%%%%%%%%%%%%%%%%%%%%%%%%%%%%%%%%%%%%%%%%
\label{sec:model}
Because RNA molecules are linear chains of bases, they  can be described as a
(quenched) sequence 
$\Sequence=(\Base[i])_{i=1,\dots,L}$ of bases
$\Base[i]\in\{\mathrm{A,C,G,U}\}$, where $L$ is the length of the sequence. 
Within this single stranded molecule some bases can pair and build a secondary
structure. 
Typically Watson-Crick base pairs, \ie A-U and C-G have the strongest
affinity to each other, they are also called
  complementary base pairs. 
Each base can be paired at most once.
For a given sequence $\Sequence$ of bases the secondary structure can be
described by a set $\Structure$ of pairs $(i,j)$ (with the convention $1\leq
i<j\leq L$), meaning that bases \Base\ and $\Base[j]$ are paired. 
For convenience of notation we further define a Matrix $(\SM)_{i,j=1,\dots,L}$
with $\SM=1$ if $(i,j)\in\Structure$, and $\SM=0$ otherwise.  
Two restriction are used: 
(i)
Here we exclude so called \emph{pseudo-knots}, that means, for any
$(i,j),(i',j')\in\Structure$, either $i<j<i'<j'$ or $i<i'<j'<j$ must hold, 
\ie we follow the notion of pseudo knots being more an element of the
tertiary structure \cite{TB99}.

(ii) Between two paired bases a minimum distance is required: $|j-i|\geq s$ is
required, granting some flexibility of the molecule (here $s=2$).

%%%%%%%%%%%%%%%%%%%%%%%%%%%%%%%%%%%
\subsection{Energy models}
\label{sec:energy_models}
Every secondary structure \Structure\ is assigned a certain energy
$E(\Structure)$; note that this energy in general depends on the \Sequence\ as
well, so it is more precisely to write $E(\Structure,\Sequence)$, but we assume
that the structure also includes the information about the sequence. 
With such an energy model it is possible to calculate the canonical partition function $Z$
of a given sequence \Sequence\ by summing over all possible structures
$Z=\sum_\Structure \E^{-\beta E(\Structure)}$, but it is  computationally
more efficient to compute it by using the partition functions of the subsequences, \ie by a dynamic programming approach.

Motivated by the observation that the secondary structure is due to building
of numerous base pairs where every pair of bases is bound by hydrogen
bonds, one assigns each pair $(i,j)$ a
certain energy $e(\Base,\Base[j])$ depending only on the kind of bases.
The total energy is the sum over all pairs
\begin{equation}
  \label{eq:paired_energy_model}
E_p(\Structure)=\sum_{(i,j)\in\Structure}e(\Base,\Base[j])\,,
\end{equation}
\eg by choosing $e(\Base[], \Base[]')=+\infty$ for non-complementary bases
\Base[], $\Base[]'$ pairings of this kind are suppressed.

Another possible model is to assign an energy \StackE\ to a pair
$(i,j)\in\Structure$  iff also $(i+1,j-1)\in\Structure$. This \emph{stacking
  energy} can be motivated by the fact that a single pairing gives some gain
in the binding energy, but also reduces the entropy of the molecule, because
through this additional binding it looses some flexibility.
 Formally the total
energy of a structure can be written as
\begin{equation}
  \label{eq:stacked_energy_model}
  E_s(\Structure)=\sum_{(i,j)\in\Structure}\StackE\SM[i+1,j-1]\,,
\end{equation}
assuming that for every  pair $(i,\,j)\in\Structure$ the bases \Base\ and
\Base[j] are complementary bases.
The total number \Ssize\ of consecutive base pairs is called the
\emph{stacking size}. 
Single base pairs are not considered as stacks,
therefore $\Ssize\geq2$ for any stack. 

Both types of energy models are discussed in the literature~% 
\cite{HB03,MET*03}, but, to our knowledge, so far no one has discussed a hybrid
model. 
We examine the sum of both models at once.
\begin{equation}
  \label{eq:energy_model}
  E(\Structure)\define E_p(\Structure)+ E_s(\Structure)
\end{equation}
where the parameters \StackE\ and $e(\Base[],\Base[]')$ can be adjusted freely,
including both models discussed above. 
Here we use 
\begin{equation}
\label{eq:pair_energy}
 e(\Base[],\Base[]')=
 \begin{cases}
   \PairE& \text{if $r$ and $r'$ are compl.\ bases}\\
   +\infty& \text{otherwise}
 \end{cases}
\end{equation}
with a pair energy $\PairE\leq0$ independent of the kind of bases. 

Due to the simple structure of the energy model, \eg the energies depend 
not on the position of the bases within the sequence or whether the paired
bases include some structure elements like hairpins, the ground state is highly
degenerated~\cite{PPR00,Har01}.

%%%%%%%%%%%%%%%%%%%%%%%%%%%%%%%%%%%%%%%%%%%%%%%%%%%%%%%%%%%%%%%%%%
\subsection{Calculating the partition function}
\label{sec:algorithm}
\begin{figure}[htbp]
  \includegraphics*[width=\picturewidth]{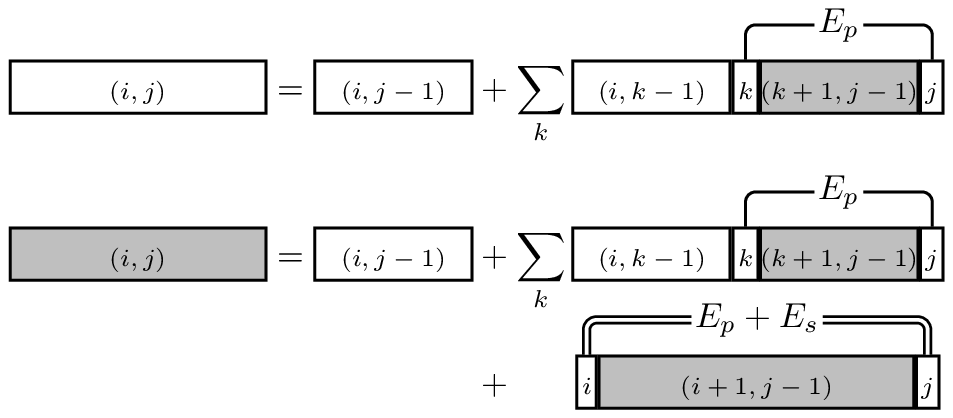}%{algorithm}
  \caption{Schematical explanation of \refeq{eq:recursion} and
    \refeq{eq:nuss_recursion}, \eg white boxes represent \Z, gray boxes \Zh.}  
  \label{fig:partition_function}
\end{figure}

Due to the fact that pseudo knots are excluded from our model (see
\refsec{sec:energy_models}), the calculation of the partition function can be 
done recursively. The algorithm is similar to that of Nussinov~%
\cite{NPG*78,DEK*98}.  
The algorithms calculates the elements of two upper-triangular matrices
$(\Z_{i,j})_{1\leq i\leq j\leq L}$ and $(\Zh_{i,j})_{1\leq i\leq j\leq L}$,
where $Z_{i,j}$ is the partition function of the
subsequence from base \Base\ to \Base[j] under the boundary condition that
bases \Base[i-1]  and \Base[j+1] are not paired, and $\Zh_{i,j}$ the partition
function under the boundary condition that bases \Base[i-1]  and \Base[j+1] are
complementary; $\Zh_{i,j}$ is only used as an auxiliary matrix. 
Then $\Z_{i,j}$ can be computed from the partition functions \Z\ and \Zh\ of
smaller subsequences in the following way 
(remember that $s$ denotes the minimum
distance between two bases of a pair)
  \begin{align}
  \label{eq:recursion}
 &\Z_{i,j}=\Z_{i,j-1}+\sum_{k=i}^{j-s-1}  
     \Z_{i,k-1}\,\E^{-e(\Base[k], \Base[j])/\kbT}\,\Zh_{k+1,j-1}\nonumber\\
    & \begin{aligned}
       \Zh_{i,j}=\Z_{i,j-1}
     +\E^{-(e(\Base[i], \Base[j])+\StackE)/\kbT}\,\Zh_{i+1,j-1}  \\
     +\sum_{k=i+1}^{j-s-1}  
     \Z_{i,k-1}\,\E^{-e(\Base[k], \Base[j])/\kbT}\,\Zh_{k+1,j-1}   
     \end{aligned}\\
      &\Z_{i,j}=\Zh_{i,j}=1\quad\text{for } i\ge j \nonumber\\
      &\Zh_{i,j}=0 \quad\text{for }0<j-i<s-2\nonumber
   \end{align}
which is schematically explained in \reffig{fig:partition_function}.
Because both matrices depend on each other, they must be calculated
simultaneously, starting along the diagonal and continuing along the
off-diagonals. 
The calculation of the partition function can be done in $\order{L^3}$ steps,
where $L$ is the length of the  sequence.
The partition function $Z$ of the entire sequence is $\Z_{1,L}$, but also the
other matrix elements are useful for generating ensembles of structures
according to the Boltzmann distribution (see
\refsec{sec:observables}). 

A similar algorithm can be used to calculate the ground state energy: 
  \begin{align}
  \label{eq:nuss_recursion}
 &\N_{i,j}=\min(\N_{i,j-1}, 
 \min_{k=i}^{j-s-1}(\N_{i,k-1}+e(\Base[k], \Base[j]))+\Nh_{k+1,j-1})\nonumber\\
    & \begin{aligned}
       \Nh_{i,j}=\min(\N_{i,j-1}, e(\Base[i], \Base[j])+\StackE+\Nh_{i+1,j-1},
       \\
     \min_{k=i+1}^{j-s-1}(  
     \N_{i,k-1}+e(\Base[k], \Base[j])+\Nh_{k+1,j-1}   ))
     \end{aligned}\\
     & \N_{i,j}=\Nh_{i,j}=0\quad\text{for } i\ge j \nonumber\\
     & \Nh_{i,j}=+\infty \quad\text{for }0<j-i<s-2\nonumber
   \end{align}
In comparison to \refeq{eq:recursion} additions are replace by
$\min$-operations, multiplications by additions and the exponentials of the
energies by the energies them self.

%%%%%%%%%%%%%%%%%%%%%%%%%%%%
\section{Observables}
%%%%%%%%%%%%%%%%%%%%%%%%%%%%
\label{sec:observables}
After calculating the partition function \Z\ for a given random sequence, we
 want to measure some quantities to compare the members of the ensemble. 
In principle most quantities could be calculated by a similar dynamic
programming algorithm introduced above, but in general the running time
behaviour would be of higher order (than three) in the sequence
length. For this reason we use a different method, where an ensemble of
structures is generated due to its Boltzmann weight~\cite{Hig96}. 
The procedure to build a sequence is essentially a backtracking algorithm:
Starting with the entire sequence a partner for an outermost base, \eg base
$L$, is chosen with the appropriate probability,  \eg base $k$, after this the
procedure is applied to the subsequences 1 to $k-1$ and $k+1$ to $L$. 
If base $L$ is chosen not to be paired at all, one uses the sequence shortened
by base $L$.
 Considering
a subsequence from base $k$ to $l$, the 
probability $p_{i,j;k,l}$ that bases $i$ and $j$ ($k\leq i<j\leq l$) are paired is given by
%
%\begin{widetext}
  \begin{equation}
    \label{eq:probability}
    p_{i,j;k,l}=
    \begin{cases}
      \begin{aligned}
  \Z_{k,l}^{-1}\,&\Z_{k,i-1}\E^{-(e(\Base[k], \Base[j])+E_0)/\kbT}\Zh_{j+1,l} \\
  &\; \text{bases $i-1$ and $j+1$ paired}\\ 
      \end{aligned}\\ \\
      \begin{aligned}
      \Z_{k,l}^{-1}\,&\Z_{k,i-1}\E^{
-e(\Base[k], \Base[j])/\kbT}\Zh_{j+1,l} \\
      &\;\text{bases $i-1$ and $j+1$ unpaired}\\
      \end{aligned}
    \end{cases}
  \end{equation}
%\end{widetext}

 For each
member of this ensemble \Ensemble\ the quantity  of interest $X$ is calculated
and the average
$\average{X}=\frac1{|\Ensemble|}\sum_{\Structure}X(\Structure)$ is used as
an approximation to the expectation value of the Gibbs-Boltzmann-ensemble; 
for large enough ensembles \Ensemble\ this average approaches the true expectation value. 

A simple observable is the energy $E$ and its fluctuations $(\Delta E)^2$, the
later is directly connected with the specific heat $c_V=(\Delta E)^2/L\,\kb
T^2$. 

Of particular importance is the overlap \overlap\ between two structures
$\Structure$ and $\Structure'$
\begin{equation}
  \label{eq:overlap}
  \overlap(\Structure,\,\Structure')\define
  \frac2L\sum_{1\leq i<j\leq L}\SM\,\SM'
\end{equation}
that is the number of bases paired to the same base in both structures
normalized such that \overlap\ lies always between zero and one. This is a
measure of how similar two structures are. 
Note however, that with this definition the overlap of one structure with
itself is $\overlap(\Structure,\,\Structure)\lneqq1$ unless all bases are
paired, where it is equal to one. 
A definition of $\overlap$ where also bases unpaired in both structures are
counted is used in~\cite{Hig96} resulting in an overlap definition that is
normalized, however this similarity measurement has the drawback that the less
bases are paired the more two structures get similar.
We further remark that for any two structures \Structure, $\Structure'$ the Cauchy-Schwarz inequality $(\overlap(\Structure, \Structure'))^2\leq
\overlap(\Structure, \Structure)\,\overlap(\Structure', \Structure')$ holds.
With this quantity \overlap\ two ensembles $\Ensemble$ and $\Ensemble'$ can be
compared, \eg looking at the distribution of
$\overlap(\Structure,\,\Structure')$ of all
$\Structure\in\Ensemble,\Structure'\in\Ensemble'$.

The ensemble averages $\average{\cdot}_{\Ensemble}$ in general depend on the
chosen sequence, therefore a further averaging over several (random) sequences
is required. This sequence average is denoted by $\qaverage{\cdot}$.
We again approximate this average by summing over a finite set of sequences.

Because of this two stage averaging, it is probably preferable instead of
looking  at $\qaverage{\average\overlap}$ directly to use
functions of the first and higher moments of \overlap, \eg the
Binder cumulant \cite{Bin81,BY85,BY88}, which  is defined by
\begin{equation}
  \label{eq:binder_cumulant}
  \BC\define\frac12\left(3-\frac{\qaverage{\average{q^4}}}{\qaverage{\average{q^2}^2}}\right)
 \end{equation} 
where $\average{\overlap^n}$ is either the average over all pairs of one
ensemble 
\begin{equation}
  \label{eq:ensemble_overlap}
  \overlap_{\Ensemble}\define \frac1{|\Ensemble|\,(|\Ensemble|-1)} 
            \sum_{\atop{\Structure,\,\Structure'\in\Ensemble}{\Structure\neq\Structure'}}  
            \overlap(\Structure,\,\Structure')\,
\end{equation}
 or the average over all pairs of two ensembles
\begin{equation}
  \label{eq:reference_overlap}
  \overlap_{\Ensemble,\,\Ensemble'}\define \frac1{|\Ensemble|\,|\Ensemble'|} 
            \sum_{\Structure\in\Ensemble,\,\Structure'\in\Ensemble'}  
            \overlap(\Structure,\,\Structure')\,.
\end{equation}
The later choice is appropriate when one is looking for a change in the
behaviour of the models when one varies the parameters in comparison 
to a reference model, 
while the former  is the better choice, if an external parameter, \eg the
temperature, is varied.

The Binder cumulant \BC\ vanishes at high temperatures, while for low
temperatures it approaches a finite value in the thermodynamic limit.

A similar quantity has been used in~\cite{PPR00}:
\begin{align}
  \label{eq:parisi_cumulant}
  \PC&\define
  \frac{\qaverage{\qvariance^2}-\qaverage{\qvariance}^2}%
       {\qaverage{\qvariance}^2}\\
\intertext{where}
\label{eq:qvariance}
\qvariance&\define L\,\left(\average{\overlap^2}-\average{\overlap}^2\right)
\end{align}
is  the variance of the $\overlap$ distribution.
This parameter \PC\ measures how the probability distribution of $\overlap$
varies from sequence to sequence. A value close to zero indicates a
self-averaging behaviour. 

%%%%%%%%%%%%%%%%%%%%%%%%%%%%%%
% Results                    %
%%%%%%%%%%%%%%%%%%%%%%%%%%%%%%
\section{Numerical Results}
%%%%%%%%%%%%%%%%%%%%%%%%%%%%%%
\label{sec:results}

In order to find some possible differences in the behaviour of the energy
model \refeq{eq:energy_model} for different parameters \PairE, \StackE,
 as introduced in \refeq{eq:pair_energy} and \refeq{eq:stacked_energy_model},
respectively, 
we
numerically calculated the quantities mentioned in \refsec{sec:observables}
above. 
In all our examples we averaged over randomly generated sequences
$\Sequence=(\Base)$,  where the probability for a specific base \Base\ at
position $i$ is $\frac14$ for all base types independent of the other bases
\Base[j\neq i].
The size of the sequence varied from $L=128$ up to $L=1024$, for the disorder
average 2000 up to 6000 random sequences were used. 
Pairing of bases are only allowed for complementary bases and the minimum
distance between to bases was chosen as $s=2$.

In \refsec{sec:basic_observables} and \ref{sec:binder_cumulant}, we vary
continuously the energy parameters between the two extreme cases
$(\PairE=0,\StackE=-1)$ and $(\PairE=-1,\StackE=0)$.
In \refsec{sec:basic_observables} we shortly discuss the averages of the
stacking size \Ssize\ and of the \overlap\ for different energy parameters.
In \refsec{sec:binder_cumulant} we examine the cross overlap
$\overlap_{\Ensemble,\,\Ensemble^\text{ref}}$, 
see \refeq{eq:reference_overlap},
between a reference ensemble $\Ensemble^\text{ref}$ generated for fixed
energy parameters, and  ensembles generated for different energy parameters.
In the following \refsec{sec:temperature_dependence} we look at the
temperature variation of various quantities without using any reference
ensemble to estimate some critical parameters. 
In the last \refsec{sec:epsilon_coupling} we apply the \epsCoup
method at $T=0$ to determine 
the critical exponent $\theta$ describing the behavior of
low-lying excitations.

%%%%%%%%%%%%%%%%%%%%%%%%%%%%%%%
\subsection{Basic observables}
\begin{figure}[ht]
  \includegraphics*[width=\picturewidth]{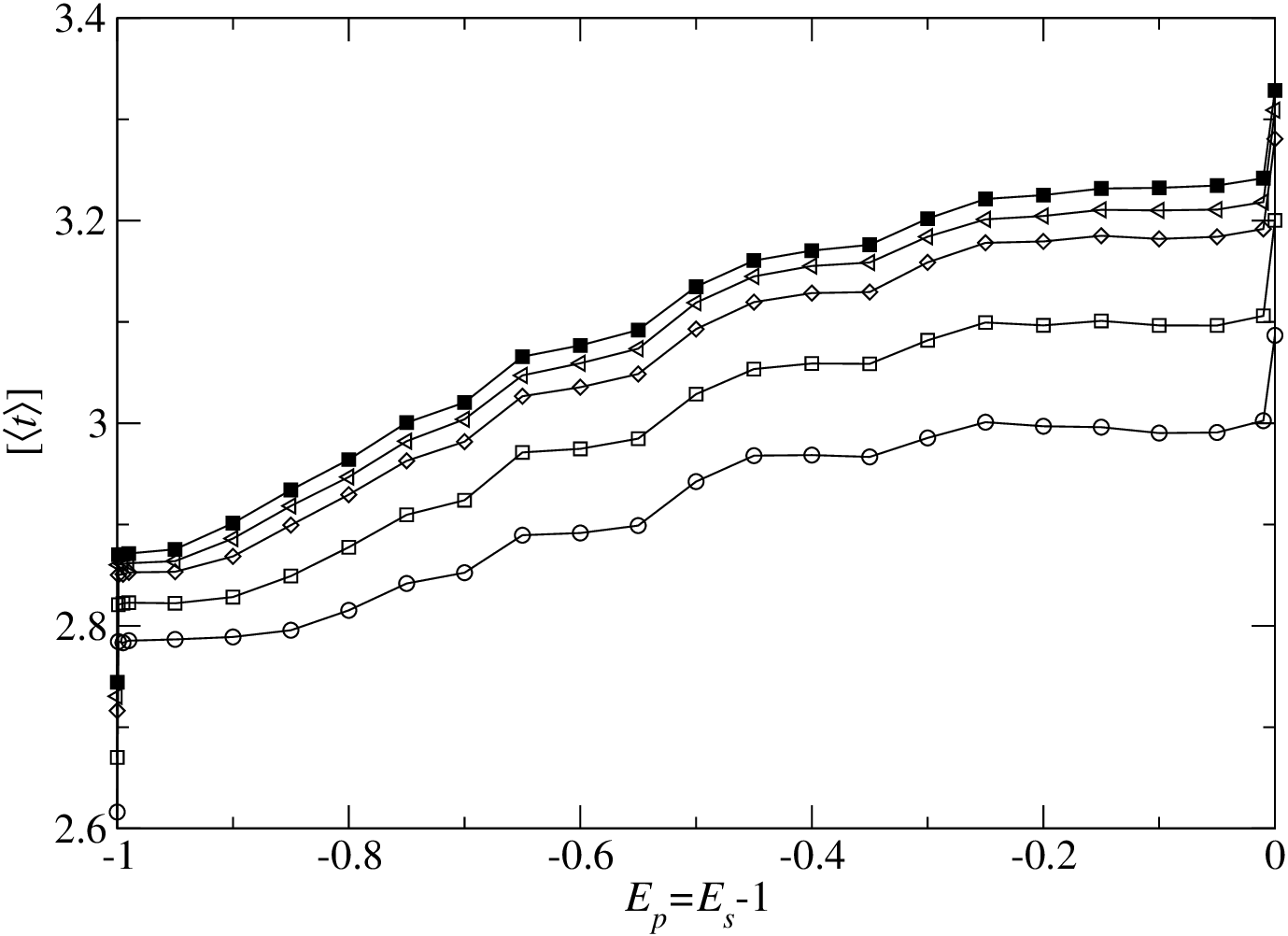}%{StackingSize}  
  \caption{ \label{fig:average_stacking_size}
     Average stacking size as a function of energy parameter $\PairE$ for
     different system sizes. At $\PairE=-1$, \ie $\StackE=0$, and $\PairE=0$
     the curves are discontinuous.
    \legendref
    \errorbars
  }
\end{figure}

%%%%%%%%%%%%%%%%%%%%%%%%%%%%%%%
\label{sec:basic_observables}
In \refsec{sec:energy_models}, where we introduced the energy model, we
opposed the pair energy to the stacking energy model. 
In \reffig{fig:average_stacking_size} we show the average size \Ssize\ of a
stacking as a function of the energy parameter $\PairE$ at temperature
$T=0$. 
We keep  $\PairE+\StackE=-1$ constant, to fix the overall
energy scale. 
For all fixed energy parameters the average stacking
size \Ssize\ increases with increasing system size,
which is expected as with increasing length the probability for longer stacks
increases. 
Also as expected is  the overall increase in the average stacking size with
the energy parameter $\StackE$, because the stacking energy prefers to build
stacks. 
The large increase of \Ssize\ while changing $\StackE$ from zero  to a nonzero
value can be explained as following:
The ground state for $\PairE=-1.0, \StackE=0.0$ is highly degenerated, while
changing to a nonzero \StackE\ only those states stay  ground states which
have a high stacking contribution to the energy. This selection increases the
average of \Ssize.
A similar argument applies at the other end, where \PairE\ changes from a nonzero value to zero.

\begin{figure}[ht]
  \includegraphics*[width=\picturewidth]{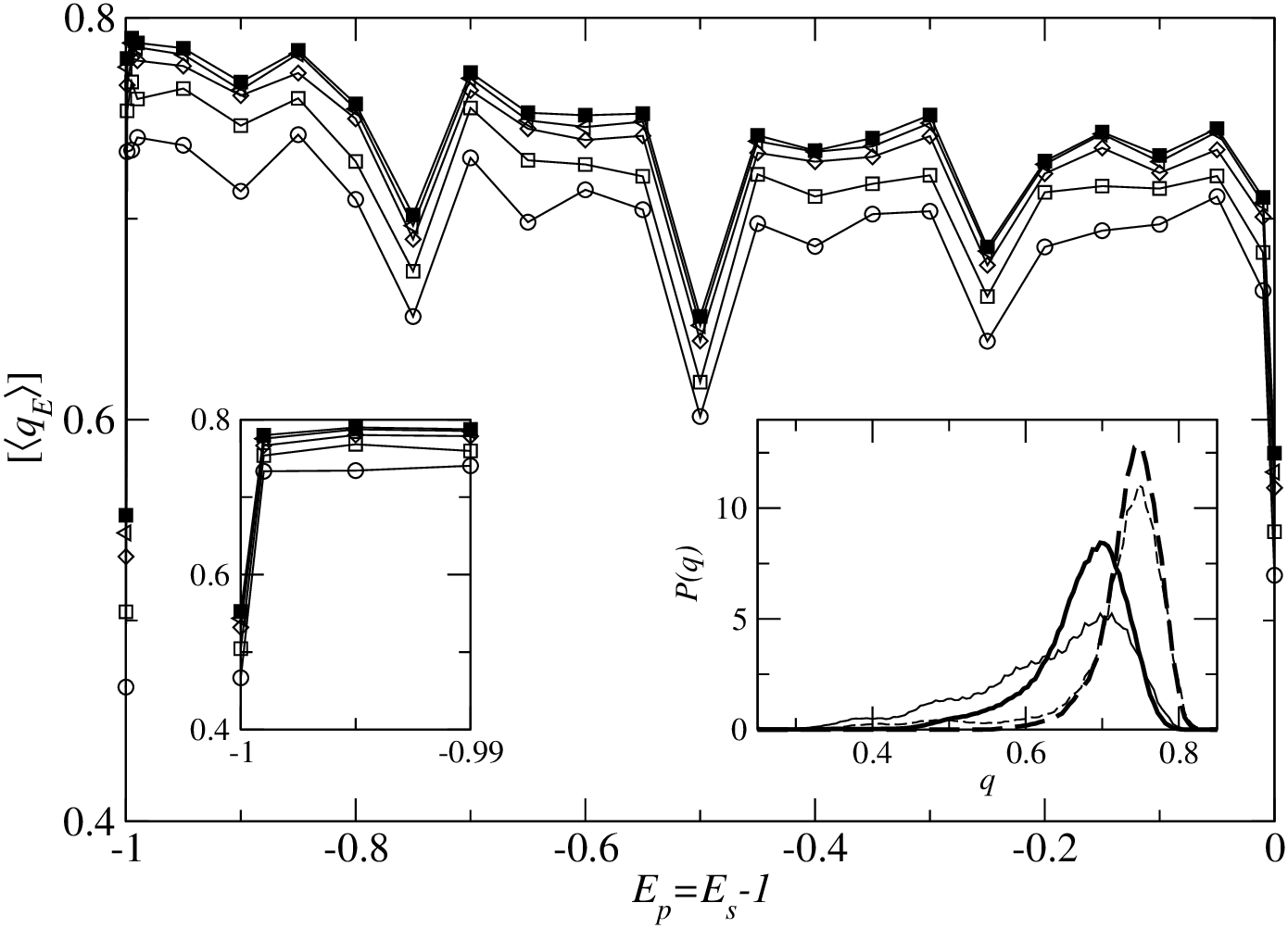}%{Overlaps}  
  \caption{  \label{fig:average_overlap}
Average overlap $\overlap_\Ensemble$ as a function of the  energy
    parameter \PairE\ for different system sizes. The local minima at
    $\PairE=-0.75,-0.5,-0.25$ are due to the commensurability of \PairE\ and
    \StackE\, and indicate a broad distribution of the ground states in
    configuration space.
    \legendref
    The left inset is an enlargement to show the discontinuity.
    The right inset is the $\overlap$-distribution for $\StackE=\PairE=-0.5$
    (solid lines) and $\StackE=-0.49, \PairE=-0.51$ (broken lines) for
    sequence lengths $L=512$ (thin lines) and $L=1024$ (thick lines).
  }
\end{figure}

\begin{figure}[ht]
  \includegraphics*[width=\picturewidth]{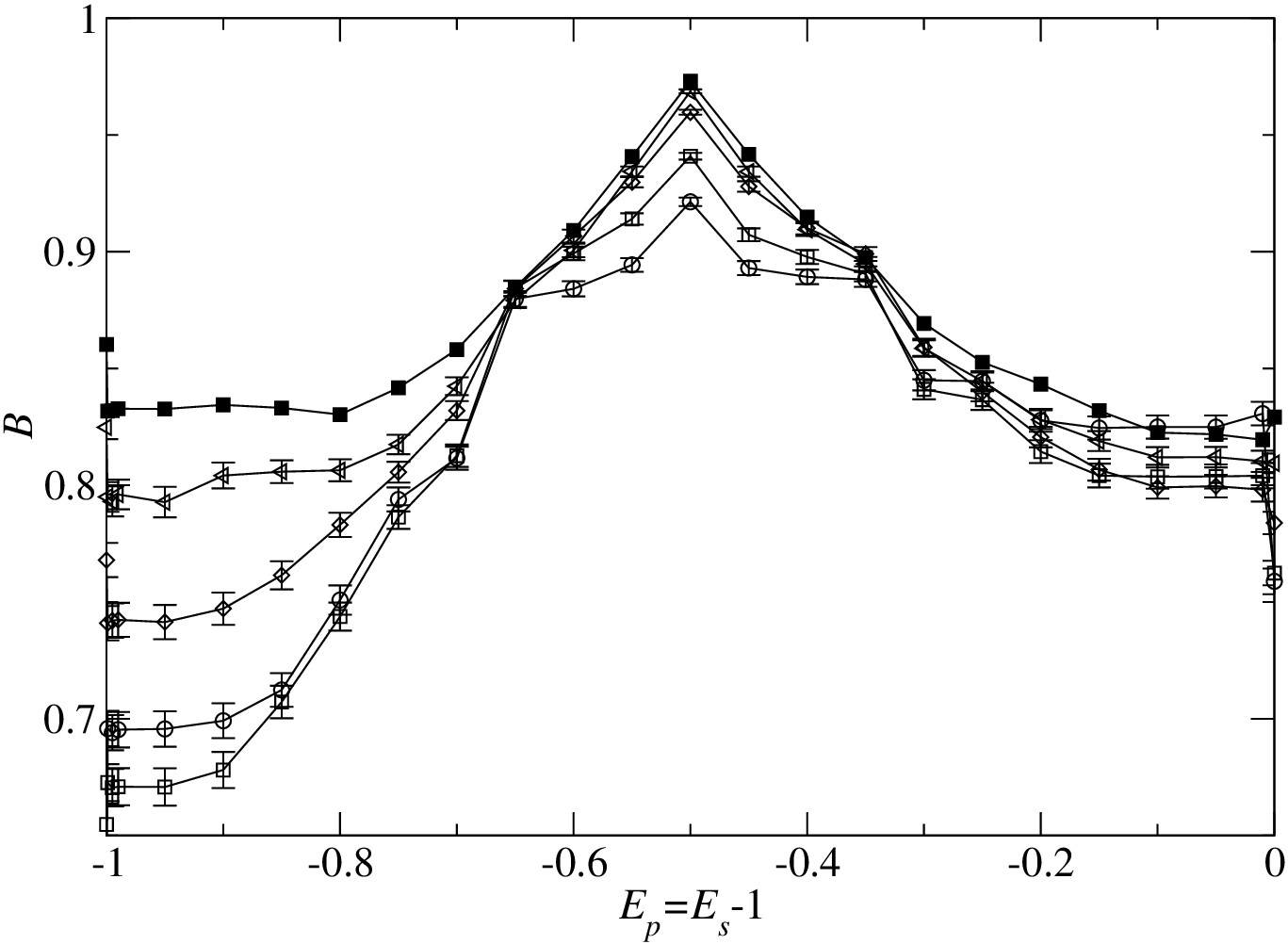}%{BinderCross}
  \caption{  \label{fig:binder_cross_cumulant}
    The Binder Cumulant \BC\ of \refeq{eq:binder_cumulant} with
    $q=\overlap_{\Ensemble,\Ensemble'}$ (see \refeq{eq:ensemble_overlap})  at
    temperature  $T=0$.  
  \legendref}
\end{figure}

Similarly the overlap \overlap\ jumps to a larger value when changing from
$\PairE=0$ to a nonzero value (\reffig{fig:average_overlap}).
In addition at positions, where $\PairE/\StackE$ are fractions with small
numerator and denominator, \eg $\frac11$ or $\frac13$,  
for this energy parameters  the ground states in configuration space are more
broadly distributed than for slightly different parameters. This can be seen
in the right inset of \reffig{fig:average_overlap}, where $q$-Distribution in
the symmetric case ($\StackE=\PairE=-0.5$) is broader than in the slightly
asymmetric case ($\StackE=-0.49, \PairE=-0.51$).
The width of this minima in the main plot, as well as that in
\reffig{fig:binder_cumulant} and \reffig{fig:binder_cross_cumulant}, is below
$\Delta E=0.001$, as one can estimate from the left inset of \reffig{fig:average_overlap}. 

For both the average stacking size and the average overlap, the
behavior changes smoothly when moving from one model to the other,
with the exception of the highly degenerate points, where we can
observe the jumps in the overlap $\overlap$. Hence, there is no sign
of a transition in between the two extremal models. To confirm this, we
next study the Binder cumulant.

%%%%%%%%%%%%%%%%%%%%%%%%%%%%%%%
\subsection{Binder cumulant}
%%%%%%%%%%%%%%%%%%%%%%%%%%%%%%%
\label{sec:binder_cumulant}
Since we introduced in \refeq{eq:energy_model} a whole class of energy
models depending on the pair energy $E_p$ and the stacking energy $E_s$, we
examined the behaviour of the Binder cumulant depending on this two energy
parameters and the sequence length $L$.
%We generated the reference Ensemble for fixed chosen $\PairE$ and
%$\StackE^\text{ref}$, and compare it with a ensembles for varying $E_p$ (the 
%stacking energy was kept constant at $\StackE=\StackE^\text{ref}$).
Second order phase transitions are characterized by crossing of the
Binder cumulant for different system sizes at the transition point.

\begin{figure}[ht]
  \includegraphics*[width=\picturewidth]{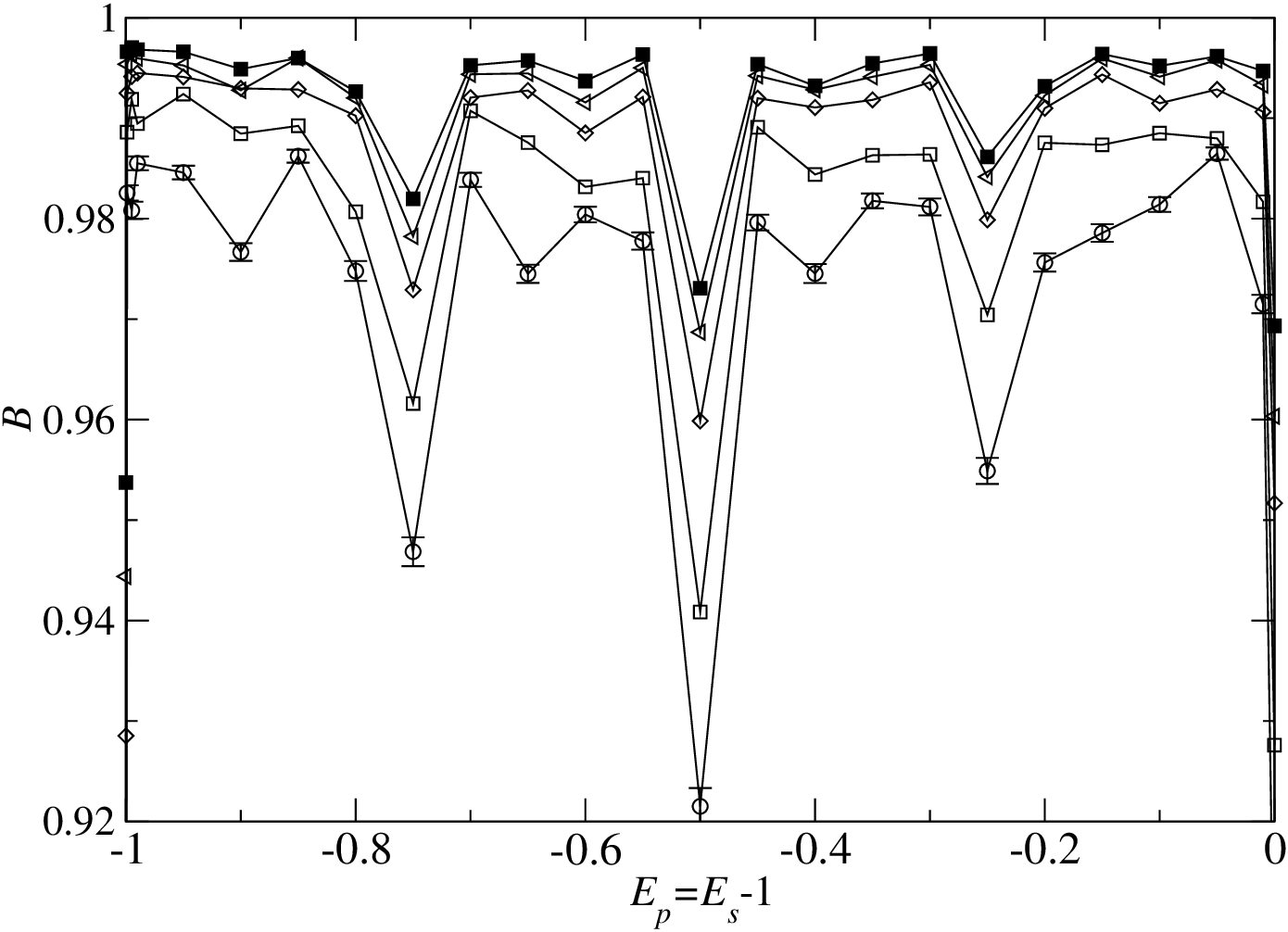}%{Binder}  
  \caption{  \label{fig:binder_cumulant}
    The Binder Cumulant \BC\ of \refeq{eq:binder_cumulant} with
    $q=\overlap_{\Ensemble}$ (see \refeq{eq:ensemble_overlap})  at temperature
    $T=0$. 
    The curves for different system sizes do not cross, and therefore \BC\
    gives no hint for a phase transition.
  \legendref
}
\end{figure}

In \reffig{fig:binder_cumulant}the Binder cumulant \refeq{eq:binder_cumulant}
is shown at $T=0$ using the ``self-overlap''
\refeq{eq:ensemble_overlap}. 
Again, the energy parameter \PairE\ and \StackE\ are varied such that always
$\PairE+\StackE=-1$ holds. 
The value of \BC\ increases with increasing system size for \PairE, \ie the
curves do not cross each other and therefore give no evidence of a phase
transition.
  The local minima are -- as the minima in \reffig{fig:average_overlap} -- due
to the commensurability of the energy parameter $\PairE$ and $\StackE$.

Further we used a reference ensemble generated for energy parameters
$\PairE=\StackE=-0.5$ and used the ``cross overlap'' of
\refeq{eq:reference_overlap}.
First, the values of \BC\ at $\PairE=\StackE=-0.5$ coincide with the values
shown in \reffig{fig:binder_cumulant}.
Similar to the observation above, \BC\ roughly increases with the
 system size, although the separation of the curves is not as clear
as in \reffig{fig:binder_cumulant}, especially in the range from $\PairE=-0.4$
to $\PairE=0.0$, where the curves coincide within the error-bars. 
To summarize, the dependence of the Binder cumulant on the energy parameters
does not indicate a phase transition.

%%%%%%%%%%%%%%%%%%%%%%%%%%%%%%%%%%%%%%%%%%%%%%%
% end: \subsection{Binder cumulant}
%%%%%%%%%%%%%%%%%%%%%%%%%%%%%%%%%%%%%%%%%%%%%%%

\subsection{Temperature dependence of the energy models}
\label{sec:temperature_dependence}
Another possible method to discriminate between the different energy
parameters is to examine the temperature dependence of some quantities,
especially the behaviour at a critical temperature. 
  In Ref.\ \onlinecite{PPR00} it was shown for a similar model that below a
critical
  temperature, 
almost all sequences fold to a compact
  structure, but for most sequences not into a single structure. 
  They point out that  this kind of
 low temperature behaviour is well known from spin
  glass and other disordered systems.
\begin{figure}[ht]
  \includegraphics*[width=\picturewidth]{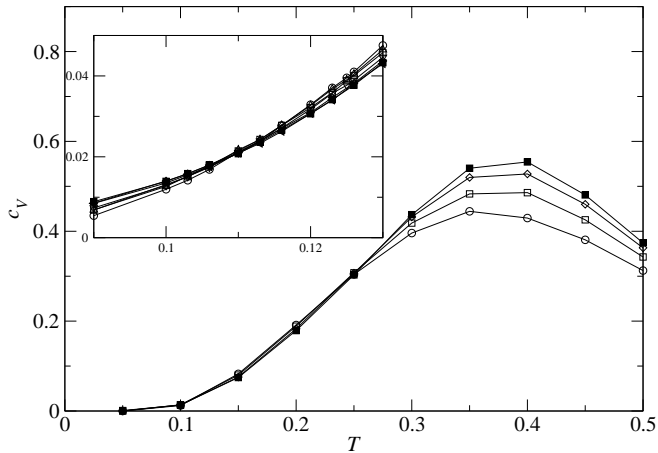}%{SpecHeat}
  \caption{  \label{fig:C_V__T}
    Specific heat $c_V$ as a function of temperature for parameters
    $\StackE=-1.0$ and $\PairE=0.0$. The curves for different system sizes
    crosses at $T\approx 0.25$ and $T\approx 0.11$. The inset is an
  enlargement of the main plot.
  \legendref}
\end{figure}
In \reffig{fig:C_V__T} we plotted the specific heat for different system sizes
as a function of temperature for $\StackE=-1.0$ and $\PairE=0.0$. 
All curves cross each other close to $T=0.11$ and $T=0.25$, which is an
evidence for a phase transition at this temperature region. 
The data for other energy parameters $(E_s, E_p)$ look similar, but the
curves cross at different temperatures.

\begin{figure}[ht]
  \includegraphics*[width=\picturewidth]{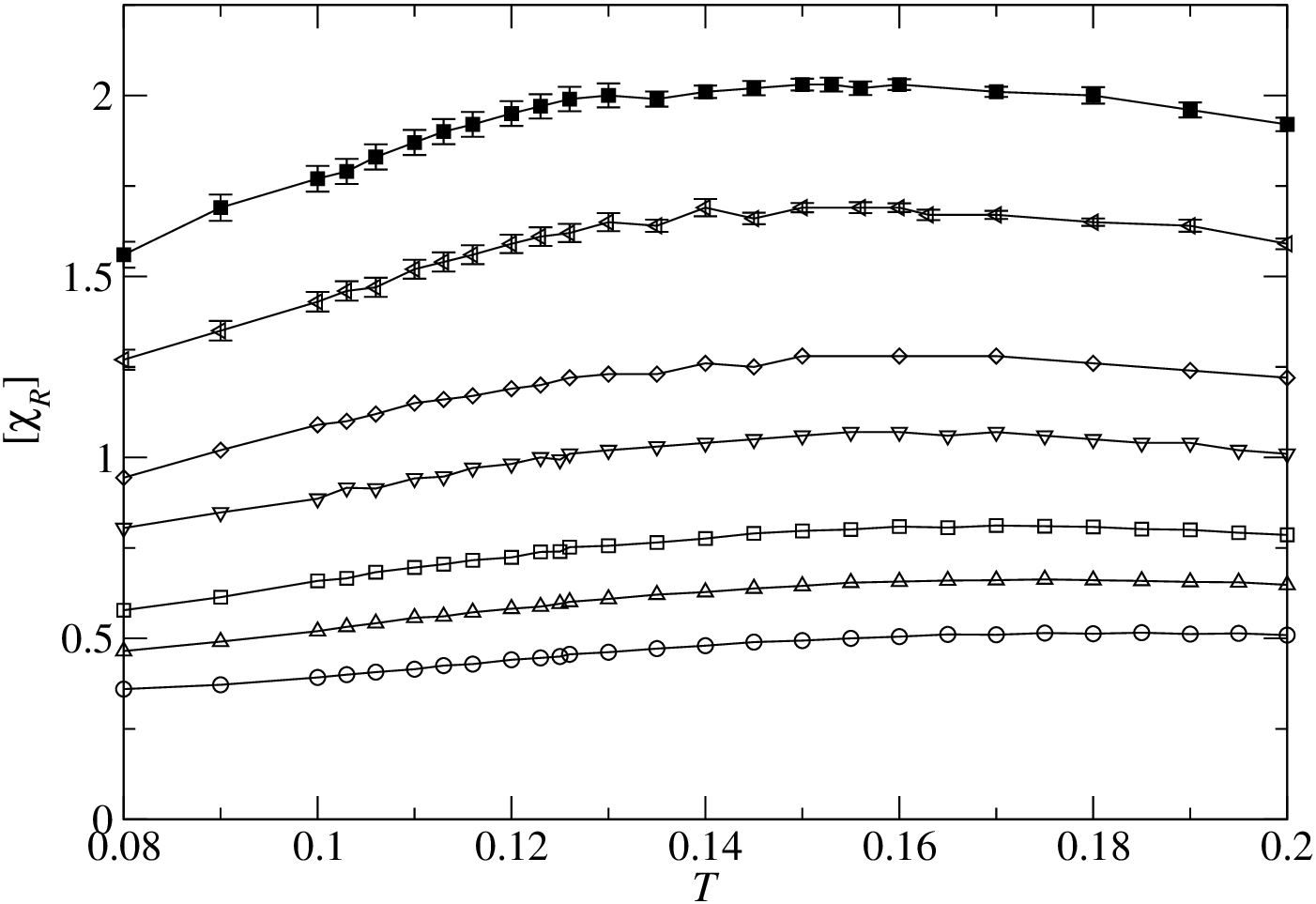}%{Xi_Es1Ep0}
  \caption{  \label{fig:qvariance_T}
    $\qvariance$ as a function of temperature for energy parameter 
    $\StackE=-1.0$, $\PairE=0.0$. 
    \legendsymbols%explanation of symbols only here
  }
\end{figure}

To determine the critical behaviour quantitatively we investigated the width
$\qvariance$ of the overlap distribution.
As can be seen from \reffig{fig:qvariance_T} all curves have a maximum and the
position of this maximum is decreasing with increasing sequence length. 
\begin{figure}[ht]
  \centering
  \includegraphics*[width=\picturewidth]{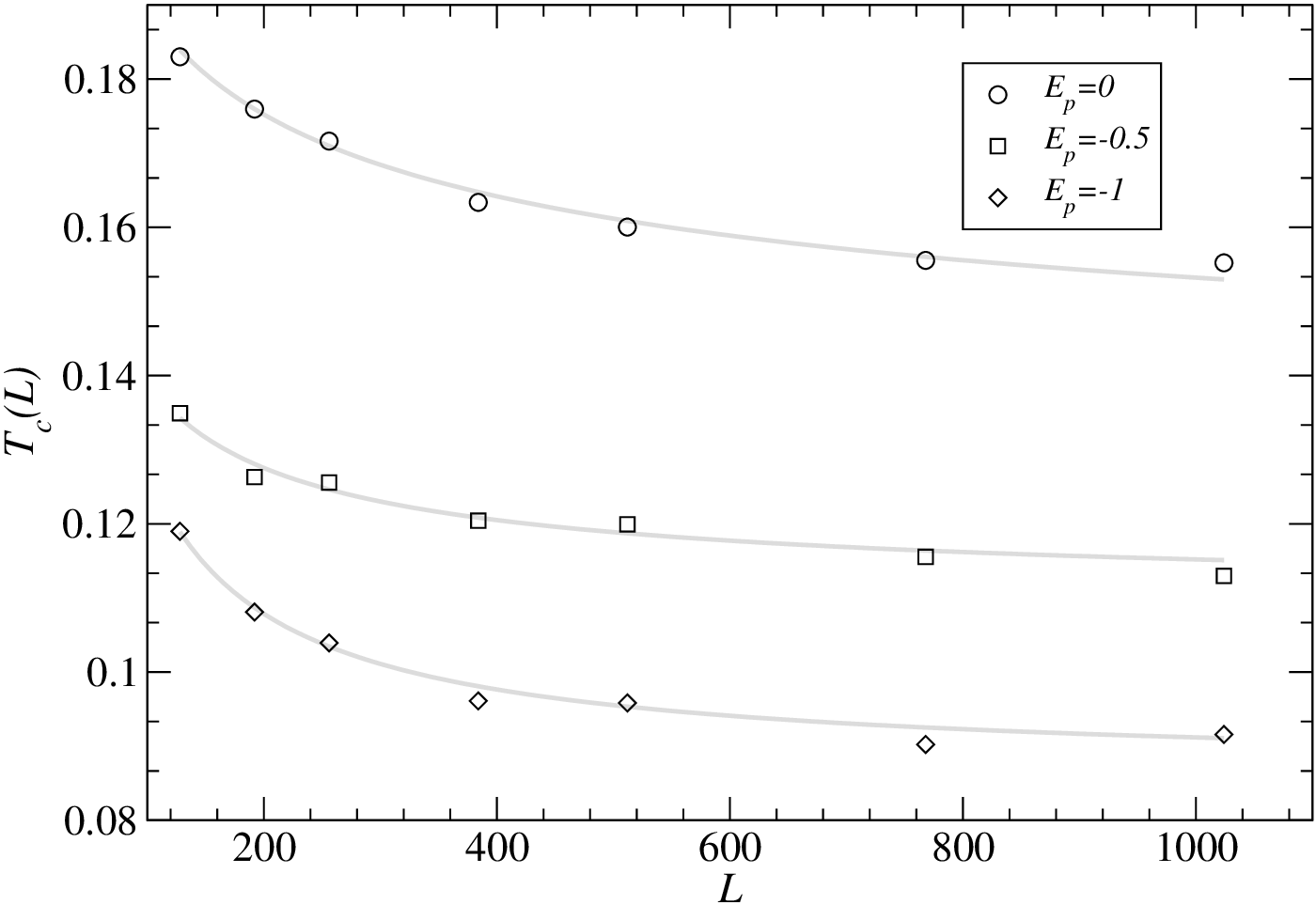}%{Fit}
  \caption{  \label{fig:scaling_fit}
    The position of the maxima of $\qvariance$ for different energy parameter. 
    The curves are fitted to the form $T_c(L)=T_c+a\,L^{-1/\nu}$, and gives
    the critical parameters of \reftab{tab:critical_parameter_chi}.
    \errorbars
  }
\end{figure}
We assume that the maximum position has the form
$T_c(L)=T_c+a\,L^{-1/\nu}$ and  fit the data to this form to get the
critical parameters (see \reffig{fig:scaling_fit}).
 The results for three different pairs of energy parameters are shown in
\reftab{tab:critical_parameter_chi}. 
\begin{table}[ht]
  \begin{tabular}{|l|l|l|l||l|}
    \hline
    \multicolumn{2}{|c|}{Energy Model} & $1/\nu$ & $T_c$& $\theta$\\
    \hline                                              % for \epsilon<0.5 
    $\StackE=0$   & $\PairE=-1$  & 0.93(15)& 0.086(3)&  0.229(38)\\
    $\StackE=-0.5$& $\PairE=-0.5$& 0.70(36)& 0.109(7)&  0.237(50)\\
    $\StackE=-1$  & $\PairE=0$   & 0.36(17)& 0.125(21)& 0.194(67) \\
    \hline
  \end{tabular}
  \caption{  \label{tab:critical_parameter_chi}
    Critical parameter of a $\qvariance$-maximum fit.
    Comparing the two limiting cases $(\StackE,\PairE)=(-1,0)$ and
    $(\StackE,\PairE)=(0,-1)$ the parameters $\nu$ and $T_c$ are different and
    indicate a quantitative change. 
    The last column belongs to the \epsCoup method in
    \refsec{sec:epsilon_coupling}.}
\end{table}
The critical exponent $\nu$ for the parameter pair $(E_s=0, E_p=-1)$ is
clearly different from that of the parameter pair $(E_s=-1, E_p=0)$, showing
that the quantitative behaviour changes when varying from one limiting case to
the other. On the other hand for the intermediate parameter set the critical
exponent is consistent with both within the error range.   

Finally, we also found (not shown) that 
 the behaviour of \PC\ of \refeq{eq:parisi_cumulant} is in agreement with
this observation and with the results~\cite{PPR00} for a two-letter RNA
model. 
For all three cases studied here,
the width of the $\overlap$--distribution varies only slightly from
realization to realization at high temperatures, while for low
temperatures the self averaging behaviour disappears.  

%%%%%%%%%%%%%%%%%%%%%%%%%%%%%%%%%%%%%%%%%%%%%%%%%%%%%%%%%%%%%%%
% end: \subsection{Temperature dependence of the energy models}
%%%%%%%%%%%%%%%%%%%%%%%%%%%%%%%%%%%%%%%%%%%%%%%%%%%%%%%%%%%%%%%

\subsection{\epsCoup}
\label{sec:epsilon_coupling}
Previously the \emph{\epsCoup method} has been used\cite{MPR02,KMM02}
to investigate low-energy excitations of RNA secondary structures.
The basic idea is to add another term to the energy model in
\refeq{eq:energy_model}, which depends on the ground state of the original
problem:
Assume $\Structure_0$ is the unique ground state of $E(\Structure)$, then a new
energy function is defined as following
\begin{equation}
  \label{eq:epsilon_energy}
  E_\varepsilon(\Structure)=E(\Structure)
             +\varepsilon\,\overlap(\Structure,\,\Structure_0)
\end{equation}
with %--- of course --- 
$\varepsilon>0$.
The additional term penalizes structures similar to $\Structure_0$, where
$\varepsilon\,\overlap(\Structure,\Structure_0)$ is largest for
$\Structure=\Structure_0$.
In general the ground state $\epsStructure$ of the new energy model
$E_\varepsilon$ will be different from $\Structure_0$. 
The difference $\Delta E(\varepsilon)\define E(\epsStructure)-E(\Structure_0)$
is an increasing function of $\varepsilon$ and $\Delta E(\varepsilon)\leq
\varepsilon$ holds.  
The later implies that $\epsStructure$ is for small enough $\varepsilon$ a low
lying excitation of the original energy model, and has the smallest overlap
with $\Structure_0$.  

The average distance $\distance(\varepsilon, L)\define
1-\overlap(\epsStructure, \Structure_0)$ between the new and the old ground
state scales as $\distance(\varepsilon, L)\propto \varepsilon\,L^{-\theta}$,
$\varepsilon$ constant,
while the average energy difference scales as $\Delta E(\distance, L)\propto
L^\theta$, $d$ constant, with the critical exponents $\theta$. For details see
Refs. \onlinecite{MPR02} or \onlinecite{KMM02}.

The assumption of a unique ground state used above does not hold for our model
used so far: in general the ground state is highly degenerated because only two
energy parameters (\StackE\ and \PairE) are used, many structures will have the
same energy. 
The degeneracy of the ground state renders the \epsCoup method as
described above almost useless. 
Since in natural RNA the contributions to the energy are more complex
different structure  will never be degenerated. 
This justifies to change the energy model slightly by adding a random
energy $\randE_{i,j}$ to the pair energies introduced in
\refeq{eq:paired_energy_model}: $e(\Base,\Base[j])\to
e(\Base,\Base[j])+\randE_{i,j}$.
There are several possibilities to choose the distributions of the $\randE$s 
(see \cite{MPR02}), here we use identical independently distributed Gaussian 
random number with zero mean $\average{\randE}=0$ and variance
$\average{\randE^2}=\randE_0^2/L$ with $\randE_0=0.1$ (the QD model in
Ref. \onlinecite{MPR02}).

\begin{figure}[ht]
\includegraphics*[width=\picturewidth]{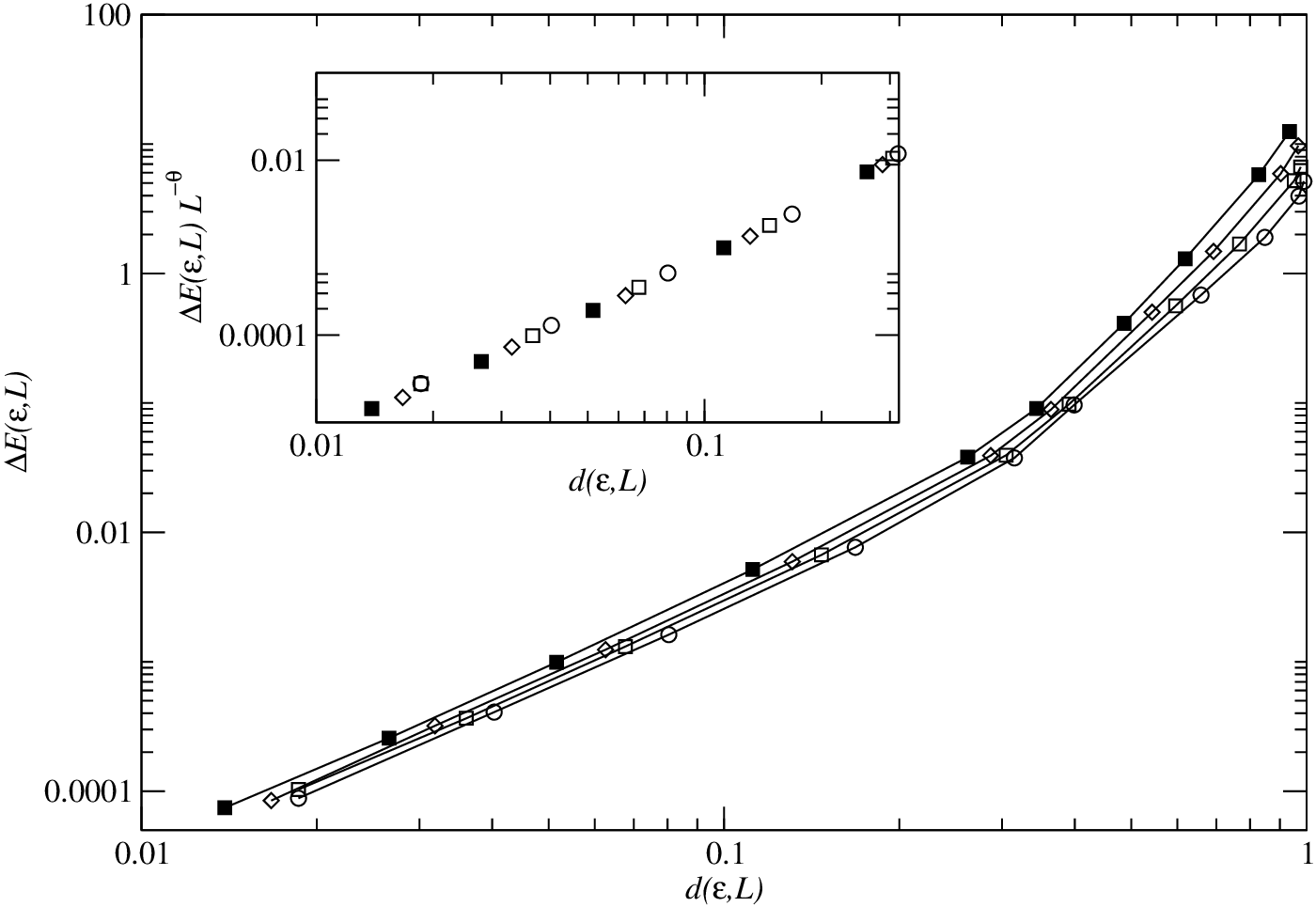}%{EpsCoupling}
  \caption{  \label{fig:epsilon_coupling}
    \epsCoup results for $\StackE=-0.5$, $\PairE=-0.5$. The energy difference
  $\Delta E(\Structure)$ between the original ground state structure
  $\Structure_0$ and the ground state structure $\epsStructure$ of the
  disturbed model is plotted as a function of the distance between this
  structures.
  The inset is a scaled plot of the data with $\varepsilon<1$ of the main
  plot. ($\theta=0.24$, see  \reftab{tab:critical_parameter_chi})
  \legendref
} 
\end{figure}

The raw result for different values of $\varepsilon \in [0.01,100]$ is
shown in  \reffig{fig:epsilon_coupling}.
A scaling plot of the
 data for $\varepsilon<1$ according the scaling form $\Delta E
 L^{-\theta}=f(\distance)$ is shown 
in the inset of  \reffig{fig:epsilon_coupling}. 
The scaling parameters $\theta$ leading to the best data collapse
for different energy parameters  are shown
in the right most column of \reftab{tab:critical_parameter_chi}. They
 are equal
within the error margins and thus does not give us a further hint of a
quantitative different behaviour for different energy parameters.
This difficulties in doing a good scaling of the data in the QD model were
already pointed out \cite{MPR02}.
However, for a different model using a scaling function
with finite-size corrections a critical exponent $\theta=0.23\pm0.05$
was obtained \cite{KMM02}, which
is close to our findings.     

%%%%%%%%%%%%%%%%%%%%%%%%%%%%%%%%%%%%%%%%%%%%%%%
% end: \subsection{$\varepsilon$-Coupling}
%%%%%%%%%%%%%%%%%%%%%%%%%%%%%%%%%%%%%%%%%%%%%%%

\section{Summary}
%%%%%%%%%%%%%%%%%%%%%%%%%%%%%%%%%%%%%%%%%%%%%%%
\label{sec:summary}
We have introduced a RNA model which continuously interpolates between two
well known models.
We sought for the answer to the question, whether there is any phase
transition of the thermodynamic behaviour.
We used both zero as well as finite temperature data.

Zero temperature results give no evidence for a phase transition apart from
trivial transitions, \eg discontinuity of \Ssize\ at points with $\StackE=0$
or $\PairE=0$. 
The curves of the Binder cumulant do not cross at a certain point, which would
be an indication of a phase transition.
Similar, the critical exponent $\theta$ derived from the \epsCoup method seems
to be independent of the energy parameters \StackE\ and \PairE, and therefore
gives no evidence for a quantitative difference in the thermodynamic
properties.
But as stated in \cite{MPR02}, the determination of the critical parameter is
rather difficult in the quasi-degenerated case studied here. 

The finite temperature results show -- in contrast to the zero temperature
data -- a quantitative dependence on the energy parameters.
The critical exponent $\nu$ for the correlation length seems to depend
on the energy model and may vary continuously while going from one
limiting model to the other.

%%%%%%%%%%%%%%%%%%%%%%%%%%%%%%%%%%%%%%%%%%%%%%%
% end: \section{Summary}
%%%%%%%%%%%%%%%%%%%%%%%%%%%%%%%%%%%%%%%%%%%%%%%

\begin{acknowledgments}
The authors have obtained financial support from the
\emph{Volkswagenstiftung} (Germany) within the program
``Nachwuchsgruppen an Universit\"aten''.
The simulations were performed at the Paderborn Center
for Parallel Computing in Germany and on a workstation
cluster at the Institut f\"ur Theoretische Physik, Universit\"at
G\"ottingen, Germany.
We thank E.~Yewande for helpful remarks.
\end{acknowledgments}

%
%%\bibliography{bbas_def,bbase}

\begin{thebibliography}{23}
\expandafter\ifx\csname natexlab\endcsname\relax\def\natexlab#1{#1}\fi
\expandafter\ifx\csname bibnamefont\endcsname\relax
  \def\bibnamefont#1{#1}\fi
\expandafter\ifx\csname bibfnamefont\endcsname\relax
  \def\bibfnamefont#1{#1}\fi
\expandafter\ifx\csname citenamefont\endcsname\relax
  \def\citenamefont#1{#1}\fi
\expandafter\ifx\csname url\endcsname\relax
  \def\url#1{\texttt{#1}}\fi
\expandafter\ifx\csname urlprefix\endcsname\relax\def\urlprefix{URL }\fi
\providecommand{\bibinfo}[2]{#2}
\providecommand{\eprint}[2][]{\url{#2}}

\bibitem[{\citenamefont{Gesteland et~al.}(1999)\citenamefont{Gesteland, Cech,
  and Atkins}}]{GCA99}
\bibinfo{editor}{\bibfnamefont{R.~F.} \bibnamefont{Gesteland}},
  \bibinfo{editor}{\bibfnamefont{T.~R.} \bibnamefont{Cech}}, \bibnamefont{and}
  \bibinfo{editor}{\bibfnamefont{J.~F.} \bibnamefont{Atkins}}, eds.,
  \emph{\bibinfo{title}{The RNA World}} (\bibinfo{publisher}{Cold Spring Harbor
  Laboratory Press}, \bibinfo{address}{New York}, \bibinfo{year}{1999}),
  \bibinfo{edition}{2nd} ed.

\bibitem[{\citenamefont{Higgs}(2000)}]{Hig00}
\bibinfo{author}{\bibfnamefont{P.~G.} \bibnamefont{Higgs}},
  \bibinfo{journal}{Quarterly Review of Biophysics}
  \textbf{\bibinfo{volume}{33}}, \bibinfo{pages}{199} (\bibinfo{year}{2000}).

\bibitem[{\citenamefont{Bundschuh and Hwa}(1999)}]{BH99}
\bibinfo{author}{\bibfnamefont{R.}~\bibnamefont{Bundschuh}} \bibnamefont{and}
  \bibinfo{author}{\bibfnamefont{T.}~\bibnamefont{Hwa}},
  \bibinfo{journal}{Phys.\ Rev.\ Lett.} \textbf{\bibinfo{volume}{83}},
  \bibinfo{pages}{1479} (\bibinfo{year}{1999}).

\bibitem[{\citenamefont{Zuker}(1989)}]{Zuk89}
\bibinfo{author}{\bibfnamefont{M.}~\bibnamefont{Zuker}},
  \bibinfo{journal}{Science} \textbf{\bibinfo{volume}{244}},
  \bibinfo{pages}{48} (\bibinfo{year}{1989}).

\bibitem[{\citenamefont{McCaskill}(1990)}]{McC90}
\bibinfo{author}{\bibfnamefont{J.~S.} \bibnamefont{McCaskill}},
  \bibinfo{journal}{Biopolymers} \textbf{\bibinfo{volume}{29}},
  \bibinfo{pages}{1105} (\bibinfo{year}{1990}).

\bibitem[{\citenamefont{Hofacker et~al.}(1994)\citenamefont{Hofacker, Fontana,
  Stadler, Bonhoeffer, Tacker, and Schuster}}]{HFS*94}
\bibinfo{author}{\bibfnamefont{I.~L.} \bibnamefont{Hofacker}},
  \bibinfo{author}{\bibfnamefont{W.}~\bibnamefont{Fontana}},
  \bibinfo{author}{\bibfnamefont{P.~F.} \bibnamefont{Stadler}},
  \bibinfo{author}{\bibfnamefont{L.~S.} \bibnamefont{Bonhoeffer}},
  \bibinfo{author}{\bibfnamefont{M.}~\bibnamefont{Tacker}}, \bibnamefont{and}
  \bibinfo{author}{\bibfnamefont{P.}~\bibnamefont{Schuster}},
  \bibinfo{journal}{Monatsh.\ Chemie} \textbf{\bibinfo{volume}{125}},
  \bibinfo{pages}{167} (\bibinfo{year}{1994}).

\bibitem[{\citenamefont{Lyngs{\o} et~al.}(1999)\citenamefont{Lyngs{\o}, Zuker,
  and Pedersen}}]{LZP99}
\bibinfo{author}{\bibfnamefont{R.}~\bibnamefont{Lyngs{\o}}},
  \bibinfo{author}{\bibfnamefont{M.}~\bibnamefont{Zuker}}, \bibnamefont{and}
  \bibinfo{author}{\bibfnamefont{C.~N.~S.} \bibnamefont{Pedersen}},
  \bibinfo{journal}{Bioinformatics} \textbf{\bibinfo{volume}{15}},
  \bibinfo{pages}{440} (\bibinfo{year}{1999}).

\bibitem[{\citenamefont{Liu and Bundschuh}(2004)}]{LB04}
\bibinfo{author}{\bibfnamefont{T.}~\bibnamefont{Liu}} \bibnamefont{and}
  \bibinfo{author}{\bibfnamefont{R.}~\bibnamefont{Bundschuh}},
  \bibinfo{journal}{Phys.\ Rev.\ E} \textbf{\bibinfo{volume}{69}},
  \bibinfo{eid}{061912} (pages~\bibinfo{numpages}{10}) (\bibinfo{year}{2004}),
  \urlprefix\url{http://link.aps.org/abstract/PRE/v69/e061912}.

\bibitem[{\citenamefont{Pagnani et~al.}(2000)\citenamefont{Pagnani, Parisi, and
  Ricci-Tersenghi}}]{PPR00}
\bibinfo{author}{\bibfnamefont{A.}~\bibnamefont{Pagnani}},
  \bibinfo{author}{\bibfnamefont{G.}~\bibnamefont{Parisi}}, \bibnamefont{and}
  \bibinfo{author}{\bibfnamefont{F.}~\bibnamefont{Ricci-Tersenghi}},
  \bibinfo{journal}{Phys.\ Rev.\ Lett.} \textbf{\bibinfo{volume}{84}},
  \bibinfo{pages}{2026} (\bibinfo{year}{2000}).

\bibitem[{\citenamefont{Higgs}(1996)}]{Hig96}
\bibinfo{author}{\bibfnamefont{P.~G.} \bibnamefont{Higgs}},
  \bibinfo{journal}{Phys.\ Rev.\ Lett.} \textbf{\bibinfo{volume}{76}},
  \bibinfo{pages}{704} (\bibinfo{year}{1996}).

\bibitem[{\citenamefont{Bundschuh and Hwa}(2002)}]{BH02}
\bibinfo{author}{\bibfnamefont{R.}~\bibnamefont{Bundschuh}} \bibnamefont{and}
  \bibinfo{author}{\bibfnamefont{T.}~\bibnamefont{Hwa}},
  \bibinfo{journal}{Phys.\ Rev.\ E} \textbf{\bibinfo{volume}{65}},
  \bibinfo{pages}{031903} (\bibinfo{year}{2002}).

\bibitem[{\citenamefont{Marinari et~al.}(2002)\citenamefont{Marinari, Pagnani,
  and Ricci-Tersenghi}}]{MPR02}
\bibinfo{author}{\bibfnamefont{E.}~\bibnamefont{Marinari}},
  \bibinfo{author}{\bibfnamefont{A.}~\bibnamefont{Pagnani}}, \bibnamefont{and}
  \bibinfo{author}{\bibfnamefont{F.}~\bibnamefont{Ricci-Tersenghi}},
  \bibinfo{journal}{Phys.\ Rev.\ E} \textbf{\bibinfo{volume}{65}},
  \bibinfo{eid}{041919} (pages~\bibinfo{numpages}{7}) (\bibinfo{year}{2002}),
  \urlprefix\url{http://link.aps.org/abstract/PRE/v65/e041919}.

\bibitem[{\citenamefont{Hartmann}(2001)}]{Har01}
\bibinfo{author}{\bibfnamefont{A.~K.} \bibnamefont{Hartmann}},
  \bibinfo{journal}{Phys.\ Rev.\ Lett.} \textbf{\bibinfo{volume}{86}},
  \bibinfo{pages}{1382} (\bibinfo{year}{2001}).

\bibitem[{\citenamefont{Liu and Bundschuh}(2003)}]{LB03}
\bibinfo{author}{\bibfnamefont{T.}~\bibnamefont{Liu}} \bibnamefont{and}
  \bibinfo{author}{\bibfnamefont{R.}~\bibnamefont{Bundschuh}},
  \emph{\bibinfo{title}{Large finite size effects in {RNA} secondary
  structures}}, \bibinfo{howpublished}{physics/0304108} (\bibinfo{year}{2003}).

\bibitem[{\citenamefont{Han and Byun}(2003)}]{HB03}
\bibinfo{author}{\bibfnamefont{K.}~\bibnamefont{Han}} \bibnamefont{and}
  \bibinfo{author}{\bibfnamefont{Y.}~\bibnamefont{Byun}},
  \bibinfo{journal}{Nucleic Acids Research} \textbf{\bibinfo{volume}{31}},
  \bibinfo{pages}{3432} (\bibinfo{year}{2003}),
  \eprint{http://nar.oupjournals.org/cgi/reprint/31/13/3432.pdf},
  \urlprefix\url{http://nar.oupjournals.org/cgi/content/abstract/31/13/3432}.

\bibitem[{\citenamefont{Mukhopadhyay et~al.}(2003)\citenamefont{Mukhopadhyay,
  Emberly, Tang, and Wingreen}}]{MET*03}
\bibinfo{author}{\bibfnamefont{R.}~\bibnamefont{Mukhopadhyay}},
  \bibinfo{author}{\bibfnamefont{E.}~\bibnamefont{Emberly}},
  \bibinfo{author}{\bibfnamefont{C.}~\bibnamefont{Tang}}, \bibnamefont{and}
  \bibinfo{author}{\bibfnamefont{N.~S.} \bibnamefont{Wingreen}},
  \bibinfo{journal}{Phys.\ Rev.\ E} \textbf{\bibinfo{volume}{68}},
  \bibinfo{pages}{041904} (\bibinfo{year}{2003}).

\bibitem[{\citenamefont{Tinoco and Bustamante}(1999)}]{TB99}
\bibinfo{author}{\bibfnamefont{I.}~\bibnamefont{Tinoco}, \bibfnamefont{Jr}}
  \bibnamefont{and}
  \bibinfo{author}{\bibfnamefont{C.}~\bibnamefont{Bustamante}},
  \bibinfo{journal}{J.\ Mol.\ Biol.} \textbf{\bibinfo{volume}{293}},
  \bibinfo{pages}{271} (\bibinfo{year}{1999}).

\bibitem[{\citenamefont{Nussinov et~al.}(1978)\citenamefont{Nussinov,
  Pieczenik, Griggs, and Kleitman}}]{NPG*78}
\bibinfo{author}{\bibfnamefont{R.}~\bibnamefont{Nussinov}},
  \bibinfo{author}{\bibfnamefont{G.}~\bibnamefont{Pieczenik}},
  \bibinfo{author}{\bibfnamefont{J.~R.} \bibnamefont{Griggs}},
  \bibnamefont{and} \bibinfo{author}{\bibfnamefont{D.~J.}
  \bibnamefont{Kleitman}}, \bibinfo{journal}{SIAM Journal of Applied
  Mathematics} \textbf{\bibinfo{volume}{35}}, \bibinfo{pages}{68}
  (\bibinfo{year}{1978}).

\bibitem[{\citenamefont{Durbin et~al.}(1998)\citenamefont{Durbin, Eddy, Krogh,
  and Mitchison}}]{DEK*98}
\bibinfo{author}{\bibfnamefont{R.}~\bibnamefont{Durbin}},
  \bibinfo{author}{\bibfnamefont{S.~R.} \bibnamefont{Eddy}},
  \bibinfo{author}{\bibfnamefont{A.}~\bibnamefont{Krogh}}, \bibnamefont{and}
  \bibinfo{author}{\bibfnamefont{G.}~\bibnamefont{Mitchison}},
  \emph{\bibinfo{title}{Biological sequence analysis}}
  (\bibinfo{publisher}{Cambridge University Press}, \bibinfo{year}{1998}).

\bibitem[{\citenamefont{Binder}(1981)}]{Bin81}
\bibinfo{author}{\bibfnamefont{K.}~\bibnamefont{Binder}}, \bibinfo{journal}{Z.\
  Phys.\ B - Condensed Matter} \textbf{\bibinfo{volume}{43}},
  \bibinfo{pages}{119} (\bibinfo{year}{1981}).

\bibitem[{\citenamefont{Bhatt and Young}(1985)}]{BY85}
\bibinfo{author}{\bibfnamefont{R.~N.} \bibnamefont{Bhatt}} \bibnamefont{and}
  \bibinfo{author}{\bibfnamefont{A.~P.} \bibnamefont{Young}},
  \bibinfo{journal}{Phys.\ Rev.\ Lett.} \textbf{\bibinfo{volume}{54}},
  \bibinfo{pages}{924} (\bibinfo{year}{1985}).

\bibitem[{\citenamefont{Bhatt and Young}(1988)}]{BY88}
\bibinfo{author}{\bibfnamefont{R.~N.} \bibnamefont{Bhatt}} \bibnamefont{and}
  \bibinfo{author}{\bibfnamefont{A.~P.} \bibnamefont{Young}},
  \bibinfo{journal}{Phys.\ Rev.\ B} \textbf{\bibinfo{volume}{37}},
  \bibinfo{pages}{5606} (\bibinfo{year}{1988}).

\bibitem[{\citenamefont{Krzakala et~al.}(2002)\citenamefont{Krzakala,
  M{\'e}zard, and M{\"u}ller}}]{KMM02}
\bibinfo{author}{\bibfnamefont{F.}~\bibnamefont{Krzakala}},
  \bibinfo{author}{\bibfnamefont{M.}~\bibnamefont{M{\'e}zard}},
  \bibnamefont{and}
  \bibinfo{author}{\bibfnamefont{M.}~\bibnamefont{M{\"u}ller}},
  \bibinfo{journal}{Europhys.\ Lett.} \textbf{\bibinfo{volume}{57}},
  \bibinfo{pages}{752} (\bibinfo{year}{2002}).

\end{thebibliography}
\providecommand{\noopsort}[1]{}

\end{document}